\documentclass[]{aa} 
\usepackage{color, soul}
\usepackage{graphics}
\usepackage{txfonts}
\usepackage{subfigure}
\usepackage{cases}
\usepackage{gensymb}
\usepackage{changepage}
\usepackage{natbib}

\begin{document}

\title{A shock near the virial radius of the Perseus Cluster}

\author{Z. Zhu\inst{1,2}
\and A. Simionescu\inst{1,2,3}
\and H. Akamatsu\inst{1}
\and X. Zhang\inst{1,2}
\and J. S. Kaastra\inst{1,2}
\and \\ J. de Plaa\inst{1}
\and O. Urban\inst{4,5,6}
\and S. W. Allen\inst{5,6,7}
\and N. Werner\inst{8}
}

\institute{SRON Netherlands Institute for Space Research, Sorbonnelaan 2, 3584 CA Utrecht, The Netherlands\\\email{Z.Zhu@sron.nl}\label{inst1}
\and  Leiden Observatory, Leiden University, Niels Bohrweg 2, 2300 RA Leiden, The Netherlands\label{inst2}
\and  Kavli Institute for the Physics and Mathematics of the Universe (WPI), The University of Tokyo, Kashiwa, Chiba 277-8583, Japan \label{inst3}
\and Trayport, Lemböckgasse 49/1A/5.OG, 1230 Vienna, Austria\label{inst4}
\and Department of Physics, Stanford University, 382 Via Pueblo Mall, Stanford, CA 94305, USA\label{inst5}
\and Kavli Institute for Particle Astrophysics and Cosmology, Stanford University, 452 Lomita Mall, Stanford, CA 94305, USA\label{inst6}
\and SLAC National Accelerator Laboratory, 2575 Sand Hill Road, Menlo Park, CA 94025, USA\label{inst7}
\and Department of Theoretical Physics and Astrophysics, Masaryk University, Kotl\'{a}rsk\'{a} 2, 61137 Brno, Czech Republic\label{inst8}
}

\abstract{
Previous X-ray studies of the Perseus Cluster, consisting of 85 {\it Suzaku} pointings along eight azimuthal directions, revealed a particularly steep decrease in the projected temperature profile near the virial radius ($\sim r_{200}$) towards the northwest (NW). 
}
{
To further explore this shock candidate, another 4 {\it Suzaku} observations on the NW edge of the Perseus Cluster have been obtained. 
These deeper data were designed to provide the best possible control of systematic uncertainties in the spectral analysis.
}
{
Using the combined {\it Suzaku} observations, we have carefully investigated this interesting region by analyzing the spectra of various annuli and extracting projected thermodynamic profiles. 
}
{
We find that the projected temperature profile shows a break near $r_{200}$, indicating a shock with $\mathcal{M} = 1.9\pm0.3$. 
Corresponding discontinuities are also found in the projected emission measure and the density profiles at the same location. 
This evidence of a shock front so far away from the cluster center is unprecedented, and may provide a first insight into the properties of large-scale virial shocks which shape the process of galaxy cluster growth.
}
{
}
\keywords {Galaxies: clusters: intracluster medium – Galaxies: clusters: individual (Perseus Cluster) – X-rays: galaxies: clusters}

\maketitle
   
\section{Introduction}
\label{sec:intro}

Approximately 80-90\% of the baryonic mass in massive galaxy clusters resides in the intracluster medium (ICM) \citep{Gonzalez2007, Gonzalez2013, Chiu2016}.
In the hierarchical large-scale structure formation model, the ICM is heated to X-ray emitting temperatures by shocks and compression as it falls into the deep gravitational potential wells dominated by dark matter \citep{Ryu2003, Molnar2009}. 
Virial shocks and accretion shocks play a key role in heating most of the baryons into a hot and diffuse state. However, neither of these phenomena have been conclusively detected yet. 
Therefore, observational probes of virial shocks constitute crucial, yet missing clues to test the current cosmological model \citep{Simionescu2021}. 

The outskirts of galaxy clusters, spanning from $r_{500}$\footnote{The radius within which the mean density is 500 times the critical density at the cluster redshift.} to the accretion shock, are the frontier of studies of the formation and evolution of the most massive haloes in the cosmic web.
A plethora of physical effects are acting in the outskirts of galaxy clusters, which closely connect with cosmic filaments in the large-scale structure.  
Compared to the inner regions of galaxy clusters, the outskirts usually have an increasingly inhomogeneous (clumpy) gas density distribution (see \citealp{Nagai2011} and  \citealp{Roncarelli2013} for simulation results), a longer electron-proton equilibration timescale, and more turbulent gas motions \citep{Nagai2013}.
Despite significant efforts over the past decade (see \citealp{Reiprich2013} and \citealp{Walker2019} for reviews), the wide outskirts region of massive galaxy clusters remains poorly explored due to their low surface brightness and the high instrumental background of orbiting X-ray satellites.
Progress in this field was driven by the low and stable particle background of the {\it Suzaku} X-ray observatory \citep{Mitsuda2007}, which allowed temperature measurements of the ICM near the virial radius using X-ray spectroscopy (see early results by e.g., \citealp{George2009}, \citealp{Bautz2009}, \citealp{Hoshino2010}, \citealp{Simionescu2011} and references thereto). 
However, its large point spread function (PSF) \citep[$\sim$2\arcmin;][]{XRT2007} largely precludes us from tracing surface brightness breaks which indicate shocks or cold fronts.
An alternative method is to take advantage of high spatial resolution X-ray observations in combination with the Sunyaev-Zeldovich (SZ, \citealp{Sunyaev1972}) imaging technique. In this context, the {\it XMM-Newton} Cluster Outskirts Project (X-COP; \citealp{Eckert2017}) offered a complementary view of the physical conditions in the outskirts of galaxy clusters \citep{Tchernin2016, Ghirardini2018}.

The Perseus Cluster (Abell 426) is the brightest cluster in the X-ray sky.
As a very popular target, Perseus has been frequently observed by various X-ray observatories, such as ROSAT, {\it Chandra}, {\it XMM-Newton} and {\it Suzaku}.
The core of the Perseus Cluster has been essential for studying the effects of active galactic nucleus (AGN) feedback, with deep {\it Chandra} data revealing ripples/sound waves and bubbles \citep{Fabian2011} as well as turbulent gas motions \citep{Zhuravleva2016} associated with the super-massive black hole (SMBH) activity.
Moreover, sloshing induced by a minor merger is also perturbing the core, forming cold fronts spanning 10-100 kpc \citep{Churazov2003, Fabian2011}.
Combining the X-ray observations from ROSAT, {\it XMM-Newton} and {\it Suzaku}, \citet{Simionescu2012} showed that these gas sloshing motions extend out to over a Mpc by measuring the thermodynamic features.
\citet{Simionescu2011} first investigated the Perseus Cluster out to its virial radius using a {\it Suzaku} mosaic extending along two different azimuths; this work was later expanded upon by \citet{Urban2014} using a total of eight azimuthal arms extending beyond {\it $r_{200}$}.
The thermodynamic properties of the ICM at large radii along the different arms have shown significant differences.
It is noteworthy that a potential shock near {\it $r_{200}$} has been identified in the northwest (NW) arm, where {\it Suzaku} spectra indicate the steepest temperature gradient among all the azimuths probed.

To further explore this potential virial shock, another four {\it Suzaku} observations on the NW edge of the Perseus Cluster have been obtained. 
Combining this unexplored {\it Suzaku} dataset and archival data, we have carefully investigated this potential shock in this work.
In Section \ref{sec:obs}, we introduce the data reduction and subsequent analysis of these {\it Suzaku} observations.
As a key part of the spectral analysis, the modeling of the cosmic and instrumental backgrounds are presented in Sections \ref{sec:cxb_model} and \ref{sec:nxb_model}, respectively. 
In Section \ref{sec:thermo_profile}, we present the projected temperature and emission measure profiles.
We further discuss the detected temperature drop and the potential shock front near the virial radius in Section \ref{sec:discussion}.
In this work, we adopt a $\Lambda$CDM cosmology with $\Omega_{m}$=0.27 and H$_{0}$ = 70 ${{\rm km}~{\rm s}^{-1}{\rm Mpc}^{-1}}$.
This gives a physical scale 1{\arcmin} =  21.8 kpc at the cluster redshift {\it z} = 0.0179 (\citealp{Struble1999}).
Our reference value for $r_{200}$ is 82{\arcmin} \citep{Simionescu2011}. 
All errors are given at the 68\% confidence level unless otherwise stated.

\section{{\it Suzaku} observations and data analysis}
\label{sec:obs}
We present the analysis of four new {\it Suzaku} pointings covering the outskirts of the Perseus Cluster in the North-West direction, named  VIR S, VIR E, VIR N and VIR W.
These pointings were taken in 2014 February, each with exposure of $\sim$25 ks. Therefore the data utilized here is approximately 2-4 times deeper than the original observation covering the NW shock front candidate and presented in \citet{Simionescu2011} and \citet{Urban2014}.
Moreover, after the NW azimuth was initially observed with {\it Suzaku}, it was realized that using a different roll angle would be preferable in order to minimize the stray light contamination in the outskirts due to the bright cool core of the cluster; the new observations make use of this favorable pointing orientation \citep{Takei2012}.

In addition to these observations, we also include a re-analysis of four old pointings (NW4, NW5, NW6, NW7) taken in 2009 August.
The observation log of all pointings is summarized in Table \ref{tab:suzaku}. 
Data from the X-ray Imaging Spectrometers (XIS) 0, 1 and 3 were analyzed and are reported here.
\begin{table}
\centering
\caption{{\it Suzaku} Observation Log}
\begin{tabular}{l c c c}
\hline\hline
Observation & OBSID & Start Time & Exposure \\
&& (UTC) & (ks) \\
VIR S & 808085010 & 2014 Feb 20 09:16:16 & 26.3 \\
VIR E & 808086010  & 2014 Feb 21 01:05:57 & 25.4 \\
VIR N & 808087010 & 2014 Feb 22 12:03:25 & 23.4 \\
VIR W & 808088010  & 2014 Feb 23 03:45:16 & 24.7 \\\hline
N4 & 804066010  & 2009 Aug 19 18:06:58 & 25.3 \\
N5 & 804067010 & 2009 Aug 20 05:55:04 & 23.6 \\
N6 & 804068010 & 2009 Aug 20 19:47:00 & 36.5 \\
N7 & 804069010 & 2009 Aug 21 15:13:02 & 36.5 \\\hline
\end{tabular}
\label{tab:suzaku}
\vspace{10pt}
\end{table}

\subsection{Data reduction}
The XIS data were analyzed following the procedure described in \citet{Simionescu2013} and \citet{Urban2014}.
In brief, we used the cleaned events files produced by the standard screening process\footnote{https://heasarc.gsfc.nasa.gov/docs/suzaku/analysis/abc/node9.html}, and applied the following additional filtering criteria:
The observation periods with low geomagnetic cut-off rigidity (COR $\leq$ 6 GV) were excluded.
For the XIS1 detector, we excluded two columns on either side of the charge-injected columns to avoid the charge leak effect.
The vignetting effect has been corrected using ray-tracing simulations of extended, spatially uniform emission.
The data reduction was performed with HEAsoft v6.26.
We have examined the 0.7-3 keV light curves of each observation with a time bin of 256 seconds, to ensure no flaring occurred during the clean exposure.
We also checked for potential contamination from solar wind charge exchange (SWCX), by plotting the proton flux measured by the WIND spacecraft’s solar wind experiment instrument\footnote{https://wind.nasa.gov/}, as shown in Figure \ref{fig:swce}. 
We found the proton flux curves of VIR S, VIR E and VIR N are very low and flat, with only a small peak at 1.2$\times$ 10$^{7}$ cm$^{-2}$ s$^{-1}$ found during the observation period of VIR W.
However, \citet{Yoshino2009} has shown that the contamination of geocoronal SWCX is negligible when the proton flux is below 4$\times$ 10$^{8}$ cm$^{-2}$ s$^{-1}$, which is much larger than the maximum flux measured during our observations.

\subsection{Imaging}
\label{sec:imaging}
We extracted images from all three XIS detectors in the 0.7-7 keV band and removed the 30$\arcsec$ region around the detector edges. 
Those pixels with an effective area less than half of the on-axis value were also masked, to minimize the influence of systematic uncertainties related to the vignetting correction.
We generated the corresponding instrumental background image from the night Earth observations.
Vignetting effects were corrected after background subtraction. 
Figure \ref{fig:im} shows the background-subtracted and vignetting-corrected mosaic of the 4 new observations, smoothed with a Gaussian kernel of 25$\arcsec$.

Combining the new mosaic with the four old observations, we visually identified 14 point sources, and exclude them using circles with radii ranging from 1$\arcmin$ to 3$\arcmin$.
{\it Suzaku} has a relatively large PSF with a half power diameter of $\sim$1.9$\arcmin$, thus we chose the minimum radius as the half power radius of the PSF.  
For brighter sources, we excluded larger regions, out to where the brightness matches the surrounding level.
More details about the excluded point sources and further check with a {\it Chandra} snapshot that covers a part of our {\it Suzaku} mosaic are discussed in the Appendix \ref{sec:psrc}.

\begin{figure}
     \centering
     \includegraphics[width=0.48\textwidth]{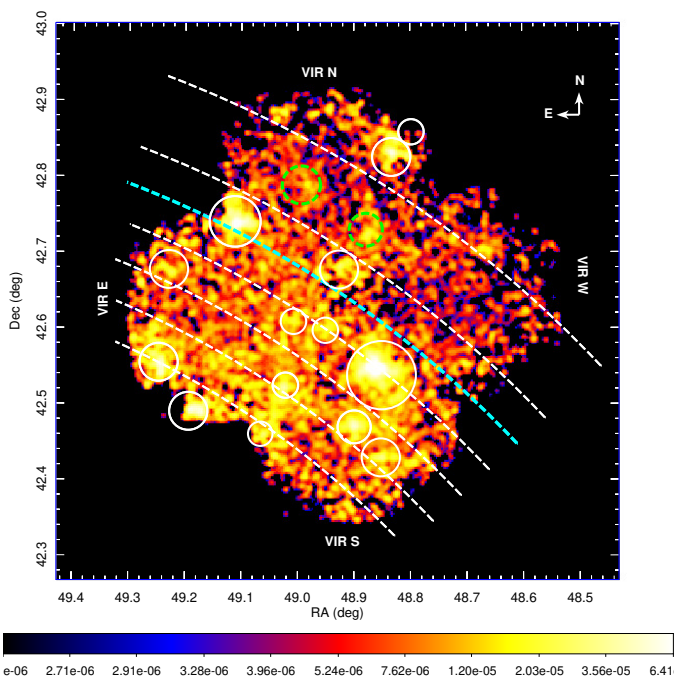}
     \caption{Exposure- and vignetting-corrected 0.7--7 keV {\it Suzaku} image obtained from the 4 new observations of the NW Perseus Cluster outskirts analyzed in this work. The identified point sources are marked with solid white circles. The cyan solid curve highlights r$_{200}$ (82$\arcmin$). The white dashed curves from bottom left to top right mark radii of 70, 73, 76, 85, and 91$\arcmin$, respectively. The image has been smoothed with a Gaussian kernel of 25$\arcsec$.  The color bar shows the surface brightness in units of counts~s$^{-1}$~arcmin$^{-2}$. 
     }
       \label{fig:im}
   \end{figure}

\subsection{Spectral analysis}
We divided the observed region from 70$\arcmin$ to 91$\arcmin$ into six partial annuli, i.e., 70$\arcmin$-73$\arcmin$, 73$\arcmin$-76$\arcmin$, 76$\arcmin$-79$\arcmin$, 79$\arcmin$-82$\arcmin$, 82$\arcmin$-85$\arcmin$, 85$\arcmin$-91$\arcmin$, and extracted spectra for each partial annulus. 
The first five annuli have a width of 3$\arcmin$ each, and are practically limited by the {\it Suzaku} PSF. 
Although the outermost annulus is slightly wider in order to improve the signal-to-noise ratio (S/N), the spatial resolution reached in this work is overall significantly better than the profile previously presented by \citet{Urban2014}. 
Redistribution matrix files (RMFs) of the XIS were produced in the standard manner using {\it xisrmfgen}, and auxillary response files (ARFs) by ray-tracing simulations using {\it xissimarfgen} (\citealp{Ishisaki2007}). 

Spectral modeling was carried out using both XSPEC \citep[v12.11,][]{Arnaud1996} and SPEX \citep[v3.06,][]{kaastra1996,kaastra2020}.
For XSPEC, the ICM was modeled as an absorbed thermal plasma in collisional ionization equilibrium using the {\it tbabs*apec} code.
The {\it apec} model is based on AtomDB v3.0.9 \citep{Foster2012}.
The spectra were grouped to have at least one count per channel.  
For SPEX, we used the {\it trafo} task to convert the OGIP format spectra and response matrices into the SPEX format.
We also modeled the ICM as an absorbed, collisionally ionized thermal plasma, i.e. {\it hot*(reds*cie)}, with SPEXACT v3.06. 
The spectra were further optimally binned (\citealp{Kaastra2016}) using the {\it obin} command.
To check the consistency, we also applied this optimal binning method via {\it ftgrouppha} in XSPEC and obtained almost identical results to the grouping method of minimum one count.
For both XSPEC and SPEX, the adopted HI column density, nH = $1.46\times 10^{21}~{\rm cm}^{-2}$, was calculated based on the Leiden/Argentine/Bonn (LAB) Survey (\citealp{Kalberla2005}).
The ICM in the Perseus Cluster is known to have a uniform metal distribution \citep{Werner2013Na}, therefore, unless noted otherwise, we fixed the metallicity of the ICM to 0.3, using the abundance table from \citet{Lodders2009}.
It's noteworthy that although a different abundance table \citep{Feldman1992} is adopted in \citet{Werner2013Na}, the differences are negligible due to a similar iron concentration in these two tables.
All spectra were fitted using the extended C-statistics (\citealp{Cash1979}).

\begin{table*}
\begin{center}
\caption{Spectral fitting models and parameters.}
\label{tab:model}
\begin{tabular}{l | c c c | c c c}
\hline\hline
Components & \multicolumn{3}{c}{XSPEC}  & \multicolumn{3}{c}{SPEX} \\
& Model & {\it kT} & Norm\tablefootmark{a} & Model & {\it kT} & Norm\tablefootmark{b} \\ \hline\hline
ICM & {\it tbabs*apec} &  -- & --   & {\it hot\tablefootmark{c}*(reds*cie)}  &  -- & --\\
LHB & {\it apec} & $ 9.25\times10^{-2} $ & $7.42\times10^{-4}$  &    {\it cie} & $9.25\times10^{-2} $ & $1.02\times10^{7}$ \\
GH & {\it tbabs*apec} & $0.138$   & $3.60 (\pm1.00)\times10^{-3}$&  {\it hot*cie} & 0.138 & $2.47\times10^{7}$  \\
HF & {\it tbabs*apec} &  $0.592 $ & $2.11 (\pm0.82)\times10^{-4}$ &  {\it hot*cie} & $0.600 $ &  $1.61\times10^{6}$ \\[5pt] \hline
& Model & $\Gamma$ & Norm & Model & $\Gamma$ & Norm\\ \hline\hline
CXB & {\it tbabs*pow} & 1.52 & -- & {\it hot*pow} & 1.52 & --\\
\hline
\end{tabular}
\end{center}
\tablefoot{The temperatures are given in keV. The column density is $1.46\times10^{21}$~cm$^{-2}$. The normalization of the CXB model components are listed in the Appendix Table \ref{tab:norm_cxb}.}
\tablefoottext{a}{defined as $\frac{10^{-14}}{4\pi[D_{A}(1+z)]^{2}}  \int n_{\rm e} n_{\rm H} dV$ in units of cm$^{-5}$, where $D_{A}$ is the angular distance and $n_{\rm e}$, $n_{\rm H}$ represent the electron and hydrogen density.} \\
\tablefoottext{b}{the emission measure $Y \equiv n_{\mathrm e} n_{\mathrm H} V$ in units of $10^{64}~\mathrm{m}^{-3}$.}  \\
\tablefoottext{c}{The electron temperature of the {\it hot} model is set to 5$\times 10^{-4}$  keV to mimic the absorption of a neutral plasma.} \\
\vspace{10pt}
\end{table*}

\subsection{X-ray foreground and background modeling}
\label{sec:cxb_model}
The X-ray foreground and background spectral model includes four components -- a power-law representing the cosmic X-ray background (CXB) due to unsolved point sources and three thermal components modeling the Galactic halo (GH, \citealp{Kuntz2000}), an additional patchy $\sim$0.6 keV hot foreground (HF, \citealp{Yoshino2009}) also due to the Milky Way, and the local hot bubble (LHB, \citealp{Sidher1996}) respectively.

Due to recent updates in the absorption models, atomic data, and abundance tables, we have re-evaluated the model parameters in order to maintain the same total flux and spectral shape as those determined by \citet{Urban2014} using a combination of {\it Suzaku} and ROSAT All-Sky Survey data.
To this end, we simulated a spectrum in XSPEC based on the old model presented in \citet{Urban2014}, $phabs*(apec+apec)+apec$, using AtomDB v2.02 and the abundance table of \citet{Feldman1992}.
In the next step, we fit this simulated spectrum in XSPEC with the current model $tbabs*(apec+apec)+apec$, adopting AtomDB v3.09 and the abundance table from \citet{Lodders2009}.
For SPEX, we fit this spectrum with $hot*(cie+cie)+cie$, using SPEXACT v3.06 and the same abundance table utilized in XSPEC.
More detailed information is listed in Table \ref{tab:model}.
Considering the $1/\sqrt{\Omega}$ dependence of the cosmic variance of unresolved point sources on the size of the extraction region $\Omega$, the systematic uncertainty of the CXB flux has increased up to $\sim$20\% due to our smaller extraction region. 
Fortunately, the ICM temperature in our region of interest is $\leq$2.5 keV, enabling us to constrain the unresolved CXB flux by directly fitting the {\it Suzaku} spectra.
Since we also have to model the Non--X-ray background (NXB; see Section \ref{sec:nxb_model}), however, this procedure needs to follow several steps. 
We first performed a fit in the 0.7-12 keV band, modeling the observations and NXB spectra in parallel, with the index and normalization of the CXB component fixed to the values adopted in \citet{Urban2014}.
After obtaining the best-fit model, then we set free both CXB parameters, freeze the NXB spectral model, and fit only the 4-7 keV band, where the CXB emission dominates.
We found the best-fit indices for each annulus are statistically consistent with the adopted value ($\Gamma = 1.52$), therefore we fixed $\gamma$ to minimize the free parameters in the subsequent analysis.  
The CXB normalization for each annulus is listed in Table \ref{tab:norm_cxb}. The best-fit CXB flux varies within a range of 8.1\%, which is smaller than the expected systematic uncertainty given the size of each extraction region.



\begin{figure*}[!tbp]
     \centering
      \includegraphics[width=0.99\textwidth]{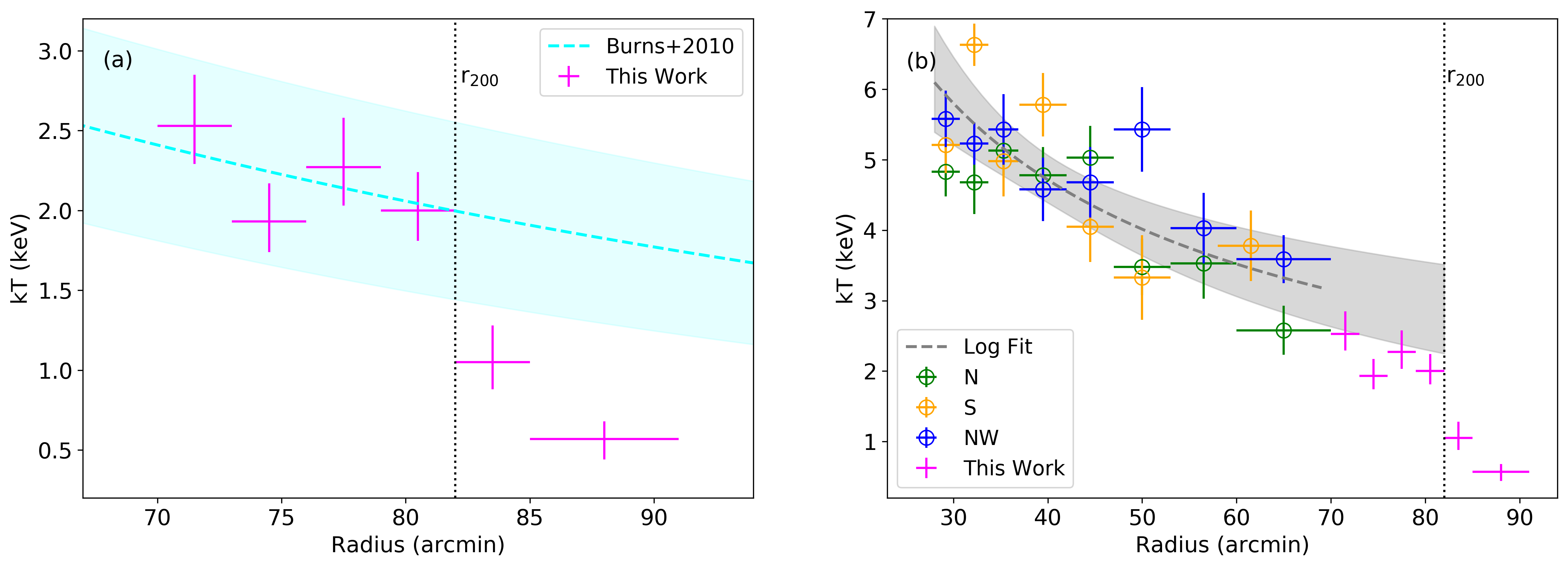}
 \caption{Projected temperature measurements (magenta points). (a) The overlaid cyan dashed line represents the prediction from the cosmological simulation (\citealp{Burns2010}). (b) We selected the measured temperatures of the three relaxed arms (i.e., NW, N and S) from \citet{Urban2014}, from 28$\arcmin$ to 70$\arcmin$, and fitted with a power-law. The grey dashed line shows the best-fit curve. All the shaded regions represent the uncertainty at 68\% confidence level. }
      \label{fig:t_profile}
  \end{figure*}

\subsection{Particle background modeling}
\label{sec:nxb_model}
The NXB spectrum of each XIS sensor was created using {\it xisnxbgen}, which extracts spectra from the integrated night-Earth observations during the period of $\pm$150 days of the Perseus outskirts observations.
We first tried to directly subtract the NXB and fit the resulting net spectra in the 0.7-7 keV band. 
However, we found that different grouping methods result in significantly different fitting results in XSPEC.
This is likely due to the fact that we are using much smaller extraction regions than employed in previous analyses of the outskirts of bright nearby clusters; hence, the NXB is noisier, which can negatively affect the results. Moreover, by construction (due to various night-Earth observations being added with different weighting factors to match the observed COR distribution), the errors on the NXB are not Poisson-distributed, which violates the conditions for using the extended C-statistic. In the high-count limit (for the NXB), this is a negligible effect; however, in this case the larger statistical errors on the NXB become important and modeling the particle background for all the three XIS detectors becomes necessary. 


The NXB is produced by charged particles and $\gamma$-rays hitting the detector from various directions.
Therefore, the spectral shape and flux of the NXB varies depending on the radiation environment of the satellite.
Due to different detector designs, the back-illuminated (BI) CCD (i.e., XIS 1) has a different NXB spectral shape compared to the front-illuminated (FI) CCD chips (i.e., XIS 0 and XIS 3) \citep{Hall2007, Tawa2008}.
As shown in Figure 1 of \citet{Tawa2008}, there are 9 instrumental lines in the 0.7-12 keV energy range.
For FI CCDs (XIS 0 and XIS 3), we modeled the NXB spectra with a power-law for the continuum and nine gaussian components for the instrumental lines. 
For BI CCD (XIS 1), we added another, broad gaussian model to account for the continuum bump above 7 keV.
We first convolved each NXB spectrum with a diagonal RMF and fitted it in XSPEC.
The line centroids of three instrumental lines (i.e., Al--k$\alpha$, Si--K$\alpha$ and Au--M$\alpha$) which are found at low energies where we expect the ICM to be detected, are set as free parameters; the centroids of the remaining lines are fixed to the values given by \citet{Tawa2008}.
We also set free all the line widths in the model and fit each spectrum.
For later analysis, the width of each individual line was fixed to the median of all spectra. 
To account for the uncertainties introduced from the normalization of the extracted NXB spectra, we added an additional constant variable as a scaling factor to our total NXB model when fitted in parallel with the source spectrum.
This scaling factor can be well constrained in the fitting energy range 0.7-12 keV, because emission from 7-12 keV is mainly contributed from the particle background. 
For spectral analysis in SPEX, we input the scaled best-fit NXB model obtained from XSPEC as a {\it file} model.
Fig. \ref{fig:nxb} show examples of the NXB spectra with their best-fit models.
\section{Thermodynamic properties}
\label{sec:thermo_profile}
The projected temperature profile obtained from the analysis described in the previous section is shown in Figure \ref{fig:t_profile}(a).
A temperature drop from 2.0$\pm$0.2 keV to 1.1$\pm$0.2 keV, moving outwards, is confirmed near {\it $r_{200}$}.
From the XSPEC normalization, we can derive the Emission Measure (EM = $\int n_{e} n_{\rm H} dV$; where $n_{\rm e}$, $n_{\rm H}$ represent the electron and hydrogen number density) via:
\begin{equation}
\centering
{\rm Norm} = \frac{10^{-14}}{4\pi[D_{\rm A}(1+z)]^{2}} EM ,
\end{equation}
\\
where $D_{A}$ is the angular distance to the Perseus Cluster.
We show the distribution of EM in Figure \ref{fig:t_em}.
Both the temperature and EM profiles show a sharp decrease between the annuli spanning 79$\arcmin$-82$\arcmin$ and 82$\arcmin$-85$\arcmin$. Comparing the measurements between these two annuli (i.e. ignoring in first instance the presence of an underlying smooth gradient), the statistical significances of these jumps are 3.4$\sigma$ and 6.3$\sigma$ for the temperature and the EM, respectively.
Assuming the plasma is fully ionized, $n_{\rm e}$: $n_{\rm H}$= 1.2 :1, we can estimate $n_{\rm e}$ from the EM.
Here, we assume the volume is a sphere of radius $r_{\rm out}$ from which we subtract an inscribed cylinder along the line of sight with radius $r_{\rm in}$, where $r_{in}$ and $r_{\rm out}$ are the inner and outer radii of each extraction region respectively.
The estimated $n_{\rm e}$ drops from $1.7\pm0.1\times10^{-4}~{\rm cm}^{-3}$ (79$\arcmin$<{\it r}<82$\arcmin$) to $1.1\pm0.1\times10^{-4}~{\rm cm}^{-3}$ (82$\arcmin$<{\it r}<85$\arcmin$).
These densities are similar to the density drop found in \citep{Walker2020}, from $1.6\pm0.1\times10^{-4}~{\rm cm}^{-3}$ to $0.8\pm0.2\times10^{-4}~{\rm cm}^{-3}$, corresponding to a cold front in the western edge of the Perseus Cluster Outskirts -- however, the temperature trend identified with $Suzaku$ suggests that at least at the NW azimuth we are looking at a discontinuity that is best described as a shock rather than a cold front.



Our analysis gives consistent measurements when including all of the available pointings or only the new observations, suggesting that the stray light contamination of the old pointings taken in 2009 was not severe.
We further compare our results with \citet{Urban2014} where only the old data are used (see the grey data points with error bars in Figure \ref{fig:t_em}), concluding that the measurements of temperatures and emission measures are consistent.
We see an improvement in statistical precision driven by fixed metallicity.
The fixed metallicity lead to the increase of the statistical precision.
This comparison also demonstrates the improvement in the spatial resolution that these new data allowed us to achieve.

\subsection{Shapes of the temperature and emission measure profiles}

In addition to the reported discontinuity in the temperature and emission measure profiles, the slope of the temperature profile appears to steepen beyond {\it $r_{200}$}. To further examine the shape of the observed temperature profile, in comparison with predicted models, we performed the following three tests:

1) We fit the projected temperature profile of the inner four data points (inside of the candidate shock front) using a power-law model (${\it kT} \propto r^{-\alpha}$),  and extrapolate this model to the pre-shock region. 
The extrapolated temperatures for the annuli 82$\arcmin$-85$\arcmin$ and 85$\arcmin$-91$\arcmin$ are ${\it kT}=1.7\pm0.4$ keV and ${\it kT}=1.5\pm0.6$ keV,  respectively. 
The gas outside of the front appears to have a lower temperature than predicted by this extrapolation; by combining in quadrature the differences between the measured and extrapolated values in the outer two annuli, we obtain a total significance of 2.1 $\sigma$.
In addition, the best fit power-law slope of the inner four and the calculated slope of outer two data points are also marginally inconsistent, with values of $\alpha = 2.6\pm1.9$ and $\alpha = 12\pm8$, respectively.

2) The inner four temperature data points match very well with the expectation from hydrodynamic cosmological simulations (cyan dashed line; \citealp{Burns2010}), while the measurements outside {\it $r_{200}$} are lower than these predictions at a cumulative 2.8 $\sigma$ confidence level, as shown in Figure \ref{fig:t_profile}(a). 
The model profile is described as 
\begin{equation}
\centering
\frac{T}{T_{\rm avg}} = A \left[1 + B\left(\frac{r}{r_{200}}\right)\right]^{\Gamma},
\end{equation}
where A = 1.74 $\pm$ 0.03, B = 0.64 $\pm$ 0.10, and $\Gamma$= -3.2 $\pm$ 0.4.
We note that any potential uncertainties in the adopted {\it $r_{200}$} value would only affect the normalization but not the slope of the predicted profile from simulations.
A higher {\it $r_{200}$} would only make the profile higher and the disagreement between the profile and outer points even more significant.

3) We fit the temperature profile of the three relaxed arms (i.e., NW, N and S) from \citet{Urban2014} with a power-law in the radial regime leading up to the region covered by the present, deeper data for the NW arm (i.e. 28$\arcmin$ $\le${\it r}$\le$70$\arcmin$). 
Although all the six data points of our work lie below this extrapolation (grey dashed line in Fig.\ref{fig:t_profile}(b)), the measurements inside {\it $r_{200}$} are on average within 1$\sigma$ from the inner trend, while the outermost two points differ at 4.1$\sigma$.
The best-fit slope of the three relaxed arms is $\alpha = 0.72\pm0.12$, which is again marginally shallower compared to the power-law trend of the outer two data points. 
 
Therefore, the outer two annuli seem to deviate from any of the three assumptions for what a continuous profile might look like, as described above.
This indicates that there is indeed a break in the temperature profile, consistent with the presence of a shock.

\subsection{Systematic uncertainties}
\label{sec:uncertainties}
\begin{figure}[tbp!]
     \centering
     \includegraphics[width=0.48\textwidth]{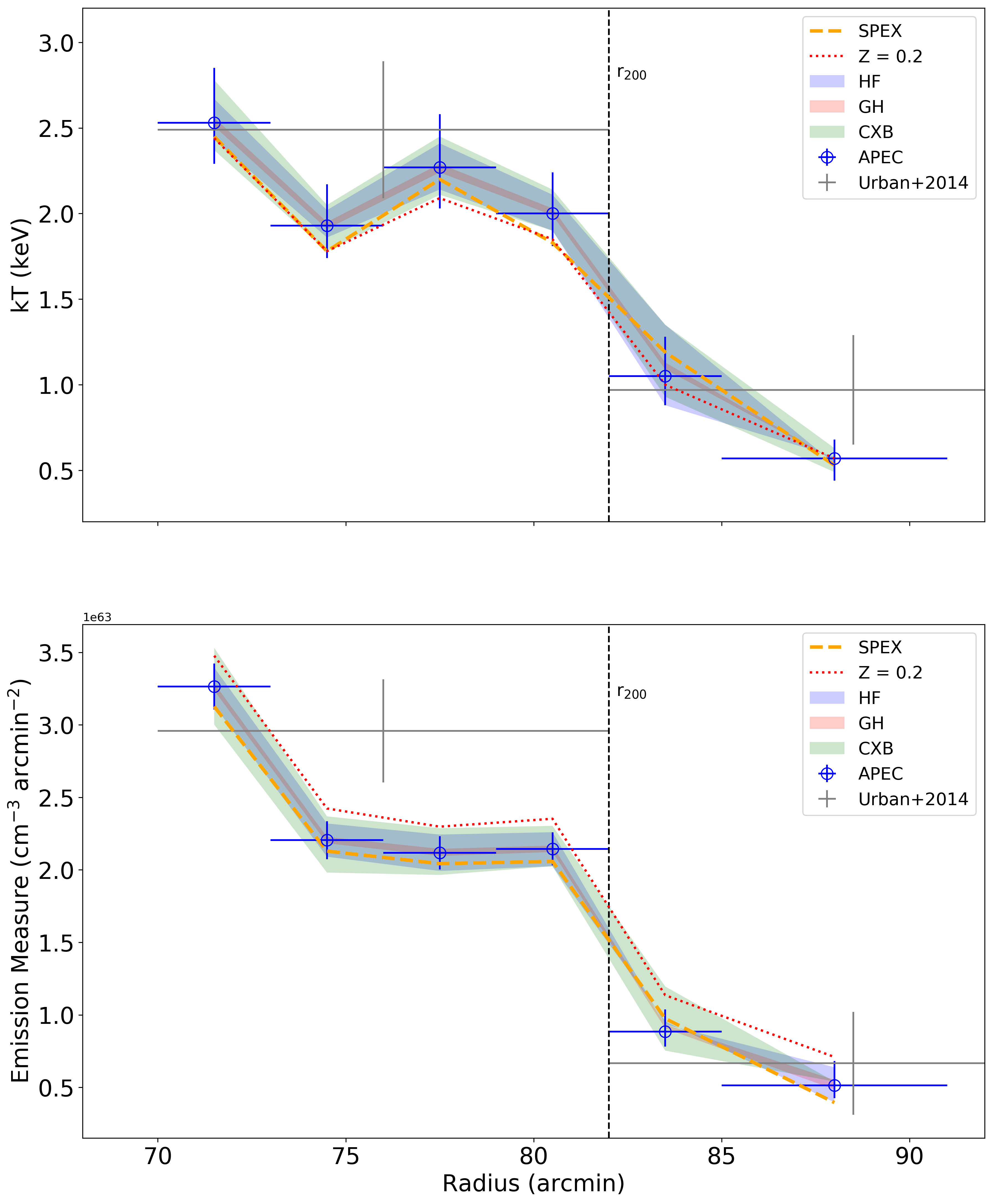}
     \caption{Profiles of the measured thermodynamic properties. Top: projected temperature profile; Bottom: the emission measure (EM = $\int n_{e} n_{\rm H} dV$) profile. The grey crosses show measurements presented in \citet{Urban2014}, using a shallower data set. The colored bands show the systematic uncertainties of the measurements due to the Milky Way foreground and the CXB (see Section \ref{sec:uncertainties} for details).  Further systematic uncertainties due to the assumed metallicity and atomic code are shown with red and orange lines, respectively.  
     }
       \label{fig:t_em}
   \end{figure}

We have carried out a thorough analysis of the systematic uncertainties related to the adopted foreground emission flux and cosmic background variance.
We adopted a 39\% systematic uncertainty for the HF and 28\% for the GH, as determined by \citet{Urban2014} and listed in Table \ref{tab:model}.
Since we allow the CXB power-law flux to be fit as a free parameter using the hard X-ray band, our measurements are not directly impacted by cosmic variance; instead, we include the statistical uncertainty on the CXB determination (see Table \ref{tab:norm_cxb} in the Appendix) as an additional source of error for the parameters measured in the full-band fits.
The colored bands in Figure \ref{fig:t_em} show that the statistical and systematic uncertainties of the measured temperatures as well as the EM are comparable.
Considering the systematic uncertainties, we obtain a temperature decrease across the putative shock front from 2.0$\pm0.2\pm0.1$~keV to 1.1$\pm0.2\pm0.2$ keV, 
i.e. the temperature drop is still present at the 98\% confidence level when both statistical and systematic errors are accounted for.
Similarly, the EM discontinuity still exists at the 3.0$\sigma$ significance level when systematic errors are added.
Furthermore, the fitting results of SPEX and XSPEC are consistent with each other, as shown by the orange dashed lines in Figure \ref{fig:t_em} that lie within the statistical uncertainty range of the blue data points.


For all the analysis in this work, we adopted a column density $N_{\rm H} = 1.46\times10^{21}~{\rm cm}^{-2}$ to keep a consistency with \citet{Urban2014}.
However, to further check the reliability of our results, we re-analysed the spectra using the total column density $N_{\rm H} = 2.10\times10^{21}~{\rm cm}^{-2}$ \citep{Willingale2013}.
The newly measured temperatures for the annuli 82$\arcmin$-85$\arcmin$ and 85$\arcmin$-91$\arcmin$ become $kT = 1.75^{+0.19}_{-0.16}$ keV and $kT = 0.88^{+0.11}_{-0.09}$ keV.
The newly derived temperature ratio between the post-shock and pre-shock region, $T_{79\arcmin-82\arcmin}/T_{82\arcmin-85\arcmin} = 1.99\pm0.30$, is consistent with the current result $1.90\pm0.40$.
Despite the uncertainty introduced from the assumed column density, the detection of a temperature drop is still reliable.
Additionally, we analyzed the IRAS 100 $\mu$m image mapping the northwest outskirts of the Perseus Cluster.
Although the central 3$\degree$$\times$ 3$\degree$ region of the Perseus Cluster has a wide range of IR emissivity (6.1--19.2 MJy/sr) \citep{Ettori1998}, this variation in the particular region of our study is small (8.1--10.3 MJy/sr; IR-<IR>$\sim$1.1 MJy/sr).
More importantly, the average IR emissivity for the post-shock arc (79$\arcmin$<{\it r}<82$\arcmin$)  and the pre-shock arc (82$\arcmin$<{\it r}<85$\arcmin$)  are 9.04 MJy/sr and 9.18 MJy/sr respectively, indicating a negligibly small variation of column density \citep{Bourdin2011}. 

Last but not the least, it is important to understand how the assumed ICM metal abundance will influence the presence and significance of the observed shock candidate.
Assuming a uniform abundance distribution, we first attempted to measure the metallicity by fitting the spectra extracted from the region spanning 70$\arcmin$<{\it r}<82$\arcmin$. 
Unfortunately, the metallicity cannot be constrained from the Fe-K line, resulting in a measurement with large error bar: {\it Z} = 0.17$^{+0.16}_{-0.12}$, and using the Fe-L complex could easily lead to large systematic uncertainties in the metallicity measurements, \citep[e.g.,][]{Buote2000, Urban2017}.
We then repeated the fits fixing the metallicity of the ICM to {\it Z}=0.2 instead of the fiducial {\it Z} = 0.3. The results are plotted in Figure \ref{fig:t_em} with red dotted lines.
The {\it kT} measurements are consistent within the statistical uncertainties, while the EM values become higher with lower metallicity. Nevertheless, the relative trends with radius remain robust regardless of the exact assumed value of {\it Z}.



\section{Discussion and conclusions}
\label{sec:discussion}
We have carefully investigated a potential shock front near the virial radius in the outskirts of the Perseus Cluster, using both archival {\it Suzaku} observations and a new $\sim$100 ks data set targeted at the region of primary interest.
The observed jump in the temperature and electron density / EM profiles presented in Section \ref{sec:thermo_profile} together indicate the presence of a shock front.
A detailed analysis of the systematic uncertainties on the measured properties (i.e., temperature and EM) further shows that this is a solid and reliable result.
We also extracted the surface brightness profile from the {\it Suzaku} mosaic and attempted to fit it with a broken power-law density model. 
Unfortunately, no significant discontinuity could be seen.
This is because the increase in line emissivity at $\sim$1 keV compensates for the lower normalization of the continuum, so that the pre-shock ($kT\sim$1 keV) and post-shock ($kT\sim$2 keV) surface brightness in the 0.7-3 keV (or even 0.7-1.2 keV) energy band is not expected to be very different.
Although it is challenging, further checks on the surface brightness profile could be realized with future {\it XMM-Newton} observations.

It is interesting to estimate the Mach number of this shock. 
To simplify the case, we assume {\it kT} and n$_{e}$ immediately outside the shock are the pre-shock conditions.
Then, we estimate the Mach number using the temperature ratio between the measurements of 79$\arcmin$-82$\arcmin$ and 82$\arcmin$-85$\arcmin$ regions.
The Mach number can be obtained by applying the Rankine-Hugoniot jump condition \citep{Landau1959}, assuming the ratio of specific heats as $\gamma$= 5/3:
\begin{equation}
\frac{T_{\rm post}}{T_{\rm pre}} = \frac{5\mathcal{M}^{4} + 14\mathcal{M}^{2} -3}{ 16\mathcal{M}^{2}} ,
\end{equation}
where T$_{\rm post}$ and T$_{\rm pre}$ are the post-shock and pre-shock temperature respectively. 
We substitute T$_{post}$/T$_{pre}$ = T$_{79\arcmin-82\arcmin}$/T$_{82\arcmin-85\arcmin} = 1.9\pm0.4$ and obtain $\mathcal{M}$ = $1.9\pm0.3$.  
This calculation does not take into account the temperature gradient that could be present in a `relaxed' configuration, however this gradient near $r_{200}$ is expected to be around 0 according to the profile of \citet{Burns2010}.
It's also noteworthy that we used projected temperatures for this calculation, therefore this Mach number could be underestimated (see Fig. 11 in \citealp{Akamatsu2017}).
We can also estimate the Mach number through the density compression factor:
\begin{equation}
C = \frac{n_{\rm e,post}}{n_{\rm e,pre}} = \frac{(\gamma+1)~\mathcal{M}^{2}}{ (\gamma-1)\mathcal{M}^{2}+2} ,
\end{equation}
which yields $\mathcal{M}=1.40\pm0.07$ when the underlying density gradient is ignored, consistent with the Mach number calculated from the temperatures within 2$\sigma$ uncertainty.  
However, unlike the temperature, we do expect a density gradient even in a relaxed case. Although it is difficult to know what this gradient would be, we can use either the data points right outside or the data points right inside the shocked annulus to estimate it. 
In that case, we get a more conservative estimate $\mathcal{M}=1.09\sim1.20$.

The location of this shock front near the virial radius of the cluster is tantalizing. 
Although virial shocks are not typically expected to be found in the inner {\it $r_{200}$} region, the shock shape can be significantly aspherical, especially in a cluster like Perseus where signs of asymmetry on large scales are present out to very large radii \citep{Simionescu2012,Walker2020}. Therefore, it is possible that the NW direction probes a special azimuth where the virial shock penetrates closer than average to the cluster core, offering a unique opportunity to detect and study such a feature.
It is noteworthy that \citet{Molnar2009} identify two different types of shocks in the outskirts of galaxy clusters, namely the "virial shock" (0.9-1.3 $r_{vir}$) and "external shock" ($\sim3 r_{vir}$).
At external shocks the infalling material gets thermalized reaching X-ray emitting temperatures for the first time, while virial shocks further heat the gas to bring it in equilibrium within the stratified entropy profile of the ICM.
The Mach number and scaled radius of the shock reported in this paper indeed lie within the ranges for "virial shock" (see Figure 4(b) in \citealp{Molnar2009}). External shocks on the other hand are much stronger (with $\mathcal{M}\sim$100) and located much further out from the cluster core.

Recently, using XMM-Newton observations towards the west of the Perseus Cluster, \citet{Walker2020} discovered a cold front $\sim$1.7 Mpc away from the core,  indicating that the gas sloshing could extend out to the virial radius.
The Perseus Cluster clearly has an active dynamic history, which opens the question of whether, instead of being a virial shock, the feature detected here could be in some way related to a past merger.
If this is the case, we can estimate the time elapsed since such a merger, $t = d / v$, where d is approximated as the radius of the shock (1.8 Mpc) and $ v = \mathcal{M}*c_{s}$.
For a pre-shock gas temperature of 1 keV,  $c_{s} = 1480*\sqrt{T/10^{8} K}~({\rm km/s})$ \citep{Sarazin1986}, yielding an age of 1.8 Gyr with the measured Mach number $\mathcal{M}$ = 1.9. 
Note that this carries a large uncertainty, since the speed of the shock will have likely not been constant over time, and there may have been a non-negligible impact parameter so that the distance of propagation need not to be equal to the shock radius.
Nevertheless, this is much shorter than the age of west cold front hinted in \citet{Walker2020} ($\sim$8.7 Gyr).
This would imply that the NW shock is not likely to be related to the same merger responsible for the W sloshing edge.
Multiple cold fronts aligned along the E-W axis are seen near the Perseus Cluster core, including one located $\sim$700 kpc to the east.
It is not clear whether all these fronts are due to the same, very old, merger, or rather to a series of consecutive mergers tracing the axis of the Perseus-Pisces supercluster.
In the latter case, one of the younger mergers in this series could indeed have produced the NW shock reported here.

Another possibility is that the W edge has a different physical origin altogether. 
\citet{Zhang2020} suggest that this cold front might be interpreted not as sloshing but as a result of the collision between the accretion shock and a ``runaway" merger shock.
In their simulation, when the runaway merger shock overtakes the accretion shock, their collision will produce three discontinuities: a rarefaction, a contact discontinuity and a merger-accelerated merger shock (MA-shock). 
According to this scenario, what we have observed might also be the runaway shock before colliding with the accretion shock considering the relatively low Mach number. 
This would mean that, along the NW direction, the runaway shock has not collided yet with the accretion shock (and is being detected in this work), while along the western direction the collision has already happened, generating the observed W contact discontinuity. 
Once again, this would imply a high degree of asymmetry in the cluster outskirts.
In short, due to the complex history and substructures in the Perseus Cluster, we cannot rule out the possibility that the NW discontinuity is a merger shock.

Previously, large-scale shocks (beyond $r_{500}$) have been investigated mostly in terms of their relation to radio relics \citep[see Table 1 in][for a summary]{Rainout2019}; even so, systematic searches rarely reveal discontinuities in both temperature and EM at the same time due to the limitation of {\it Suzaku} PSF and projection effects \citep[e.g.,][]{Akamatsu2013sys}.
One exception is found at the Coma Cluster relic, where both a steep temperature drop (from 3.6 keV to 1.5 keV) and the shape of the surface brightness profile together suggest a shock with Mach number $\mathcal{M}=2.2\pm0.5$ \citep{Akamatsu2013}. This shock is associated with the in-falling NGC4839 group and is located close to {\it $r_{200}$} of the main Coma Cluster core.
Similarly, {\it XMM-Newton} and {\it Suzaku} observations of the galaxy cluster Abell 2744 also reveal the presence of a shock front at the radio relic with $\mathcal{M}=1.7_{-0.3}^{+0.5}$, which is seen both in the density and temperature profiles, and located 1.5 Mpc east of the cluster core \citep{Eckert2016, Hattori2017}. 
On the other hand, it is extremely rare to detect shock features in the far outskirts of cool-core clusters -- this task is all the more difficult because we do not have the position of the radio relic to inform us about where to search for such shocks.
To our knowledge, the shock reported here in the NW of the Perseus Cluster is the farthermost shock detected in a cool-core cluster of galaxies, confirmed by both density and temperature measurements, and lacking a radio counterpart. Regardless of its exact nature, the detection of this shock paves the way for understanding the heating processes that are at play in the outer parts of galaxy clusters, where the ICM meets the surrounding large-scale structure.





\begin{acknowledgement}
We thank the anonymous referee for constructive suggestions that improved this paper.
ZZ, AS, HA, JSK, JdeP are supported by the Netherlands Organization for Scientific Research (NWO).
The Space Research Organization of the Netherlands (SRON) is supported financially by NWO.
XZ is supported by the the China Scholarship Council (CSC). 
SWA acknowledges support from the U.S. Department of Energy under contract number DE-AC02-76SF00515.
NW was supported by the GACR grant 21-13491X.
This research is based on observations obtained from the \emph{Suzaku} satellite, a collaborative mission between the space agencies of Japan (JAXA) and the USA (NASA). 

\end{acknowledgement}

\bibliography{ref}
\bibliographystyle{aa}

\begin{appendix}
\section{Solar Proton flux variation during {\it Suzaku} observations}

\begin{figure}
    \centering
    \includegraphics[width=0.5\textwidth]{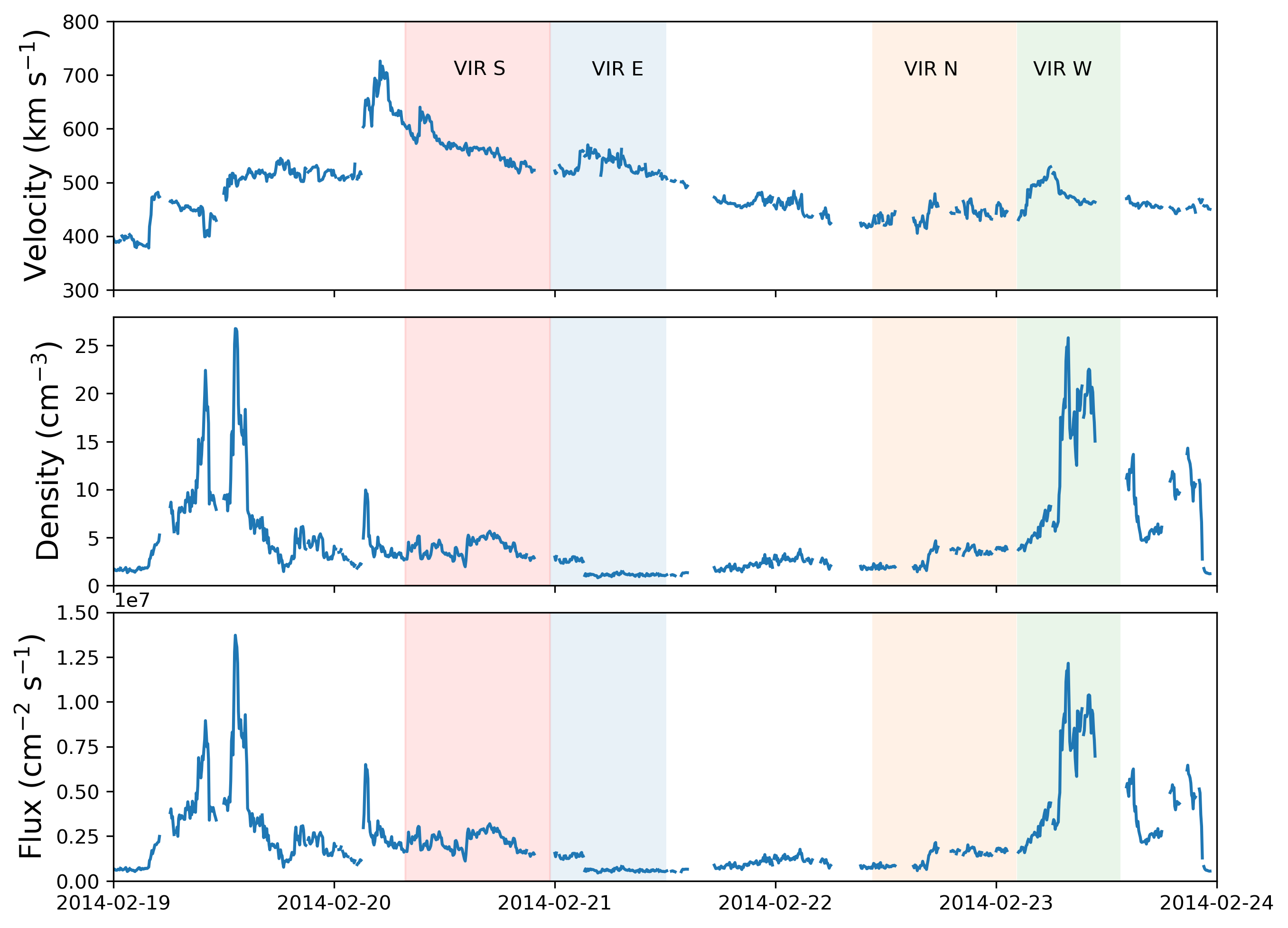}
    \caption{Solar proton flux measured by the WIND spacecraft's solar wind experiment instrument. The red, blue, yellow and green shaded regions denote the time coverage of {\it Suzaku} observations VIR S, VIR E, VIR N, VIR W, respectively.}
    \label{fig:swce}
\end{figure}

Spectral evidence of geocoronal solar-wind charge exchange (SWCX) was first obtained during a {\it Chandra} dark moon observation \citep{Wargelin2004}.
Using {\it XMM} observations of SWCX emission, \citet{Snowden2004} claimed that the enhancement of the soft X-ray intensity was correlated with solar-wind proton flux variations. 
The emission lines of C$_{VI}$, O$_{VII}$, O$_{VIII}$, Ne$_{IX}$ and Mg$_{XI}$ are detected in the enhancement.
Therefore, SWCX becomes a significant foreground contamination in the study of soft X-rays below $\sim$1 keV. 
In our case, the measured temperature of the outermost annulus is $\sim$0.5 keV, which makes it particularly important for us to check the variation of solar wind proton flux.
Here, in Fig.\ref{fig:swce} we plotted the solar proton flux measured by WIND spacecraft's solar wind experiment instrument during 2014 Feb 19-24, which covers the periods of all 4 new {\it Suzaku} observations used in our analysis.
Fig.\ref{fig:swce} shows that the light curve is low and stable in VIR S, E, N.
There is a peak in VIR W but it is more than 10 times lower than the flux where SWCX typically becomes important in observations.

\section{Point source selection}
\label{sec:psrc}
We have a $\sim$7 ks shallow {\it Chandra} observation (ObsID 20521; PI: S.Walker), which however only covers a part of our {\it Suzaku} mosaic. 
For this reason we rely on the {\it Suzaku}-only point source identification for the default analysis. 
We have also checked the 15 point sources detected by this {\it Chandra} snapshot, and most of them were already excluded.
However, two sources (marked with green dashed circles in Figure \label{fig:im}) are found in the outermost (85$\arcmin$-91$\arcmin$) annulus, that were not part of our original point source list. 
We redid the analysis for this particular annulus and while the CXB norm decreased from $1.19^{+0.04}_{-0.06}$ to $1.08^{+0.05}_{-0.05}\times10^{-3}$ due to the exclusion of these additional point sources, the cluster properties have not changed significantly ( we obtain $kT = 0.52^{+0.13}_{-0.13}$, $Norm = 0.92^{+0.40}_{-0.19}\times10^{-3}$ compared to $ kT = 0.57^{+0.11}_{-0.13}$, $Norm = 0.94^{+0.31}_{-0.16}\times10^{-3}$ for the default analysis). 
This further illustrates that our background modeling method is robust.

Furthermore, we extracted the {\it Suzaku} spectra of all selected point sources. 
Except for one of them found to be a bright star (2MASS J03164923+4229460; $V_{mag}$ = 8.76), the spectra of the rest could be well fitted by absorbed power-laws with $\Gamma \sim1.8$, which are typical of  background AGNs.

\section{Cosmic X-ray Background} 

Despite our efforts to remove detected point sources in our {\it Suzaku} field of view, the cosmic variance of unresolved point sources still introduces relatively large uncertainties to our spectral analysis.
Compared to \citealp{Urban2014}, the systematic uncertainty of the CXB flux has increased up to $\sim$20\% due to our smaller extracted annuli.
Fortunately, the ICM temperature in our region of interest is $\leq$2.5 keV, enabling us to constrain the unresolved CXB flux using 4-7 keV {\it Suzaku} spectra
The step-by-step procedure is described in Section \ref{sec:cxb_model} and the obtained CXB normalization for each annulus is listed here in Table \ref{tab:norm_cxb}. 

\begin{table}[btp!]
\renewcommand{\arraystretch}{2}
\caption{XSPEC normalizations and their statistical uncertainties of the CXB model components for different annuli, in units of $10^{-3}$ photon~cm$^{-2}$~keV$^{-1}$~s$^{-1}$ at 1 keV.  }
\label{tab:norm_cxb}
\centering
\begin{tabular}{c | c | c}
\hline\hline
Annulus & $\Gamma$ & power-law norm \\\hline
70$\arcmin$--73$\arcmin$ &  & $1.18^{+0.12}_{-0.13} $\\
73$\arcmin$--76$\arcmin$ & & $1.34^{+0.09}_{-0.07}$\\
76$\arcmin$--79$\arcmin$ & 1.52 & $1.33^{+0.07}_{-0.08} $\\
79$\arcmin$--82$\arcmin$ & &$1.19^{+0.05}_{-0.07} $\\
82$\arcmin$--85$\arcmin$ & &$1.21^{+0.07}_{-0.07} $\\
85$\arcmin$--91$\arcmin$ & & $1.19^{+0.04}_{-0.06}$\\\hline
\end{tabular}
\end{table}


\section{Particle background modeling}
\label{sec:nxb}
\begin{figure*}
 \centering
    \includegraphics[width=0.4\textwidth, angle=270]{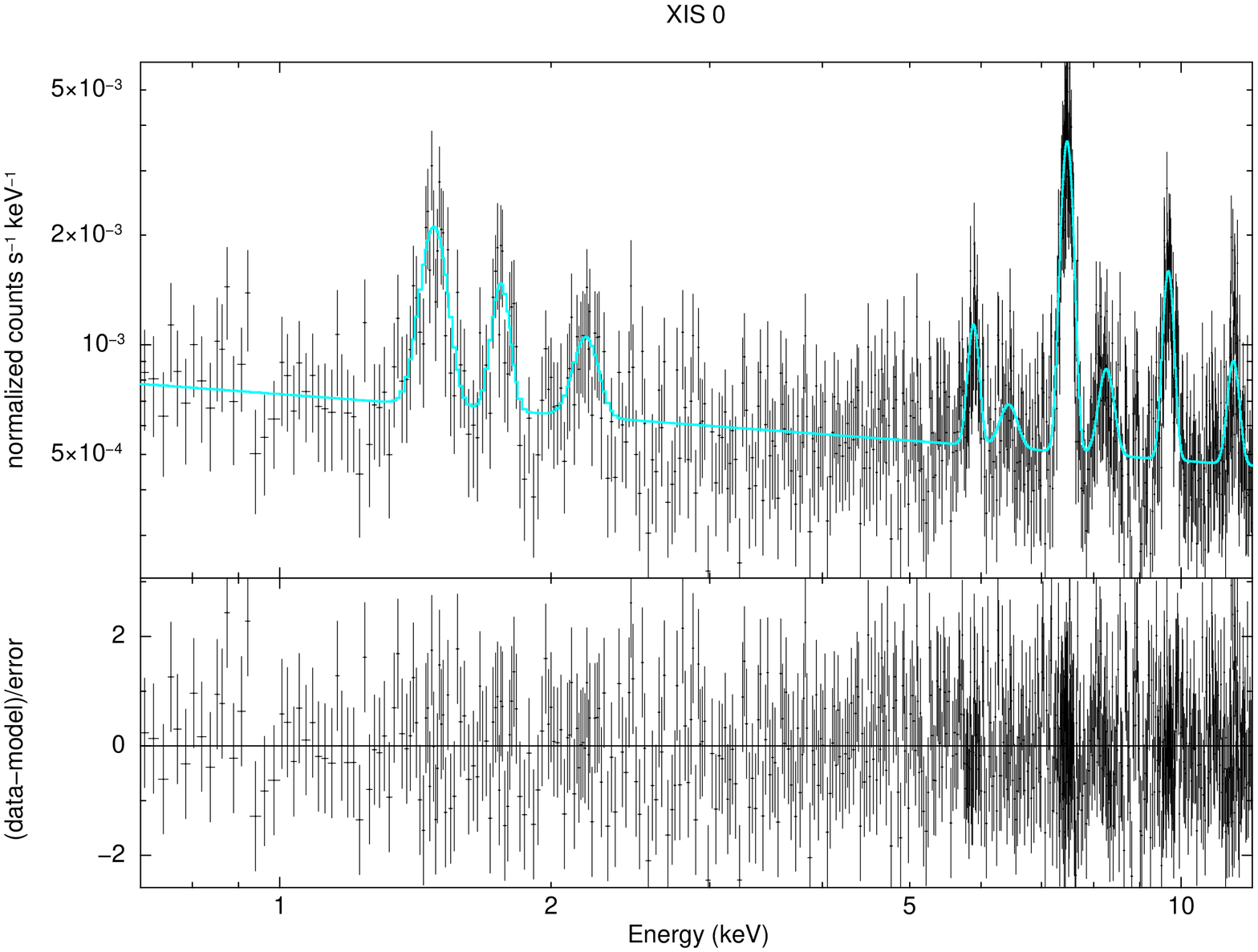}
        \includegraphics[width=0.4\textwidth, angle=270]{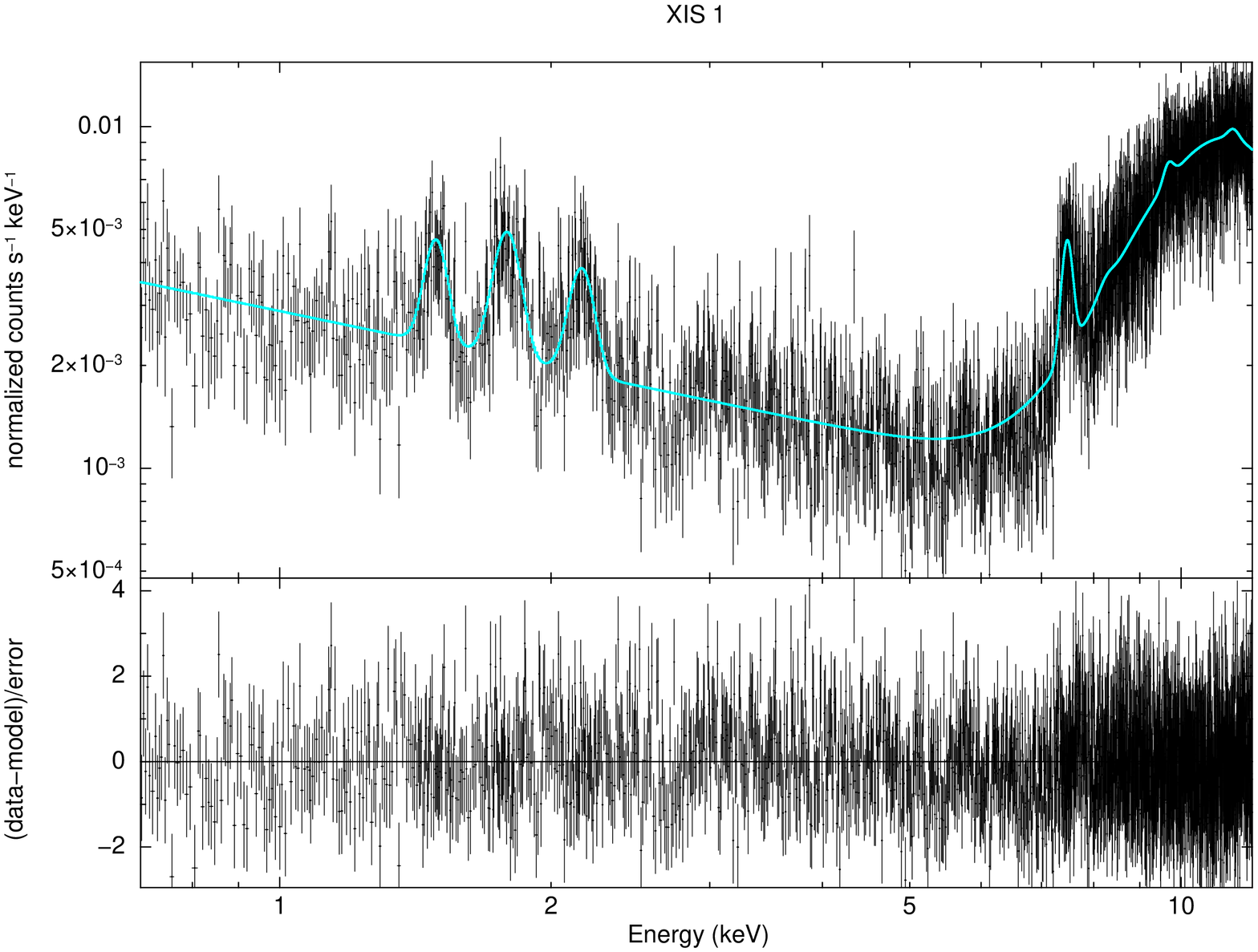}
            \includegraphics[width=0.4\textwidth, angle=270]{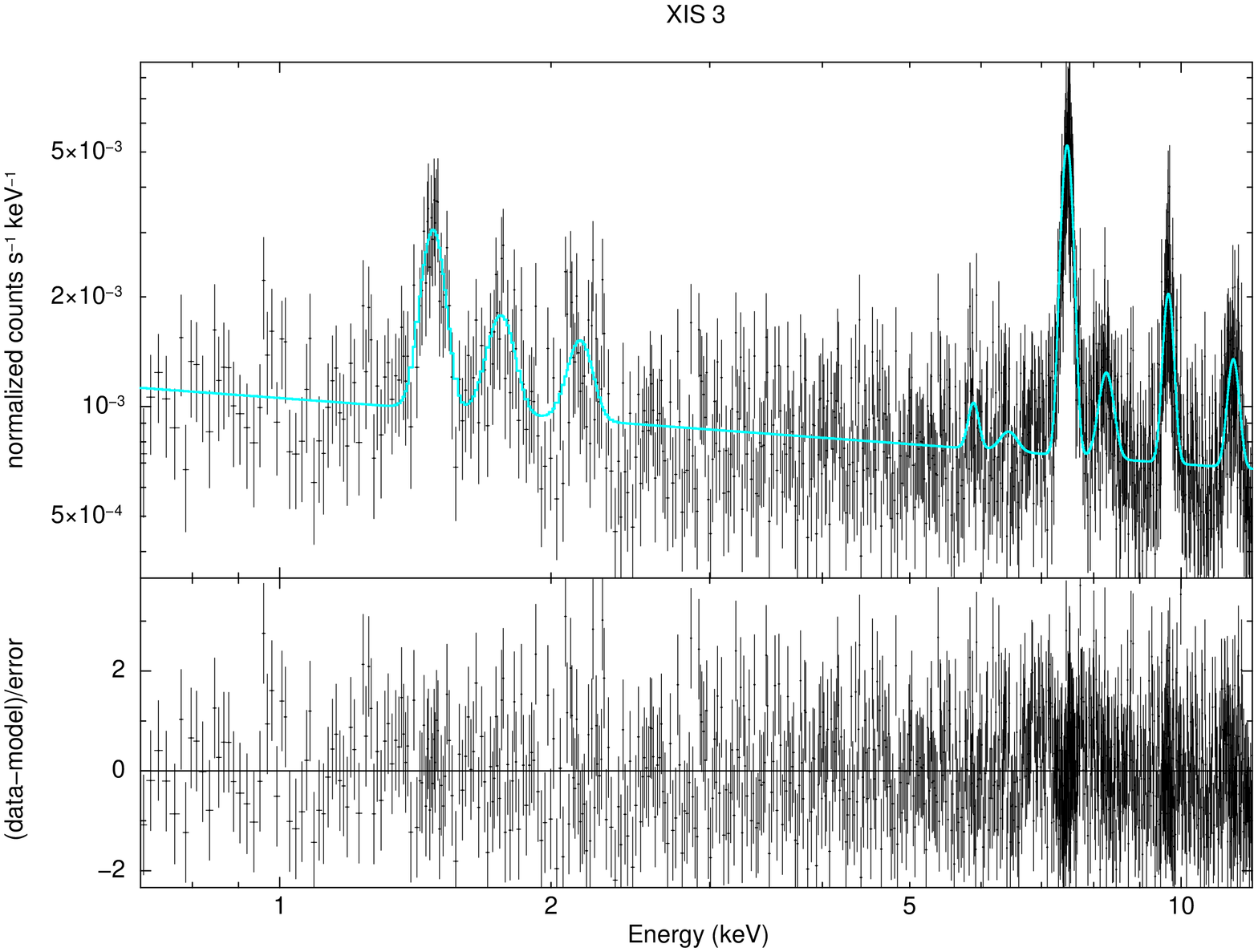}
    \caption{The typical NXB spectrum of each XIS sensor created by {\it xisnxbgen}. The best-fit models are denoted with cyan lines. The data has been rebinned to S/N$\geq$3 for plotting purpose, while the fit was performed on the data optimally binned using the {\it obin} command \citep{Kaastra2016}. }
    \label{fig:nxb}
\end{figure*}

In Fig. \ref{fig:nxb}, we show the typical NXB spectrum with its best-fit model of each XIS sensor.
The spectra for VIR S, annulus 70$\arcmin$-73$\arcmin$ are shown here as an example, and the NXB shapes are very similar for all other pointings and annuli.
More details can be found in Section \ref{sec:nxb_model}.
 
\section{Projected measurements}
In Table \ref{tab:proj}, we give the projected measurements corresponding to the data points plotted in Figure \ref{fig:t_em}.  

\begin{table}[btp!]
\renewcommand{\arraystretch}{2}
\caption{Projected temperatures, XSPEC normalizations and EM ($\int n_{e}n_{H}dV$) of the six annuli.}
\label{tab:proj}
\centering
\begin{tabular}{c c  | c c | c  | c c }
\hline\hline
\multicolumn{2}{c}{Annulus} & \multicolumn{2}{c}{$kT$} & $Norm$& \multicolumn{2}{c}{$EM$} \\\hline
$r_{in}$  & $r_{out}$ & XSPEC & SPEX & XSPEC   & XSPEC  & SPEX \\\hline
70  &   73  & $2.53^{+0.32}_{-0.24}$  &  $2.45^{+0.34}_{-0.27}$ & $5.98^{+0.29}_{-0.29}$  & $3.31_{-0.16}^{+0.16}$& $3.17^{+0.17}_{-0.16}$\\
73  &   76  & $1.93^{+0.24}_{-0.19}$ & $1.78^{+0.32}_{-0.19}$ & $4.04^{+0.24}_{-0.24}$  & $2.23_{-0.13}^{+0.13}$& $2.16^{+0.17}_{-0.14}$ \\
76  &   79  & $2.27^{+0.31}_{-0.24}$ &  $2.20^{+0.27}_{-0.35}$ &$3.88^{+0.21}_{-0.21}$  & $2.14_{-0.11}^{+0.11}$ & $2.07^{+0.14}_{-0.13}$\\
79  &   82  & $2.00^{+0.24}_{-0.19}$& $1.83^{+0.26}_{-0.15}$ & $3.93^{+0.21}_{-0.20}$  & $2.17_{-0.11}^{+0.11}$ & $2.09^{+0.12}_{-0.10}$\\
82  &   85  & $1.05^{+0.23}_{-0.17}$ & $1.19^{+0.34}_{-0.09}$ &$1.62^{+0.28}_{-0.19}$  & $0.90_{-0.10}^{+0.15}$& $0.99^{+0.24}_{-0.10}$\\
85  &   91  & $0.57^{+0.11}_{-0.13}$ & $0.53^{+0.11}_{-0.13}$ &$0.94^{+0.31}_{-0.16}$  & $0.52_{-0.09}^{+0.17}$& $0.40^{+0.16}_{-0.09}$ \\\hline
\end{tabular}
\tablefoot{The units of annuli are arcminutes and the XSPEC normalizations are given in units of $10^{-3}$ 		photon~cm$^{-2}$~keV$^{-1}$~s$^{-1}$. The {\it EM} is in units of $10^{62}~{\rm cm}^{-3}~{\rm arcmin}^{-2}$. }
\end{table}

\section{An example of the spectrum}
\begin{figure*}
 \centering
    \includegraphics[width=0.9\textwidth]{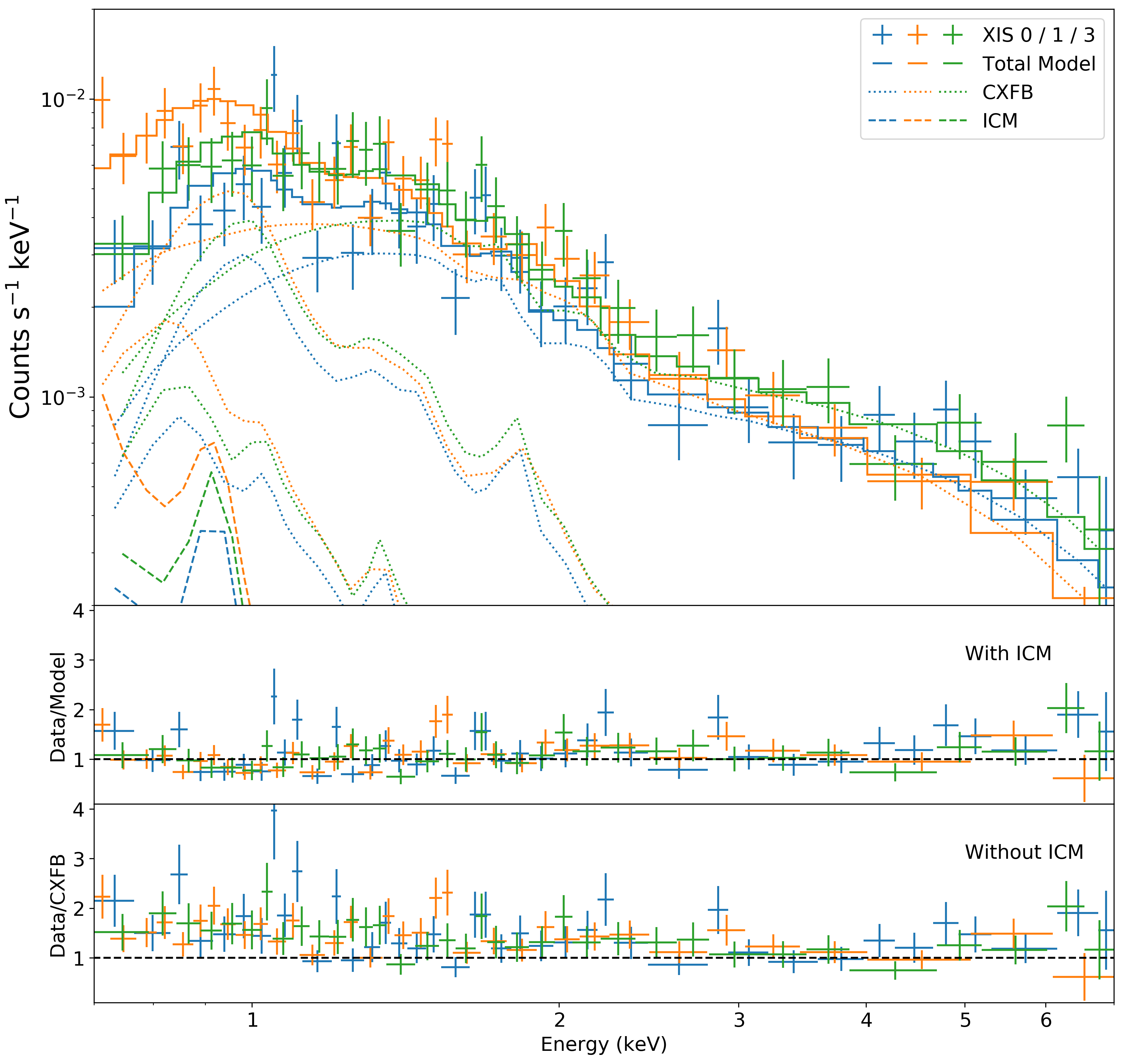}
    \caption{Example of 0.7--7 keV spectra and the best-fit models of the first annulus outside $r_{200}$ (82$\arcmin$< {\it r} < 85$\arcmin$) in the NW arm. The best-fit NXB models have been subtracted. Data from all three detectors are shown – XIS0 (blue), XIS1
(yellow) and XIS3 (green). The CXFB models are marked with dotted lines and ICM models are denoted with dashed lines. The data has been rebinned to S/N$\geq$3 for illustration purpose, while the fit was performed on the data binned with a minimum of one count. The lower two panels show the fitting residuals including and excluding the ICM component respectively.}
    \label{fig:spec_example}
\end{figure*}

In Fig. \ref{fig:spec_example}, we show the spectrum extracted from the annulus in the potential pre-shock region immediately outside $r_{200}$ (82$\arcmin$< {\it r} < 85$\arcmin$) along the NW arm.  
While in the spectral analysis, for each annulus we fit spectra from different pointings in parallel, here we show the merged spectrum to give an idea of the nice quality of our dataset.
In the bottom panel of Fig. \ref{fig:spec_example}, we plot the ratio between the observed and CXFB fluxes. 
We see a clear excess around 1 keV, indicating the significance of the ICM signal.

\end{appendix}

\end{document}